\documentclass[aps,prl,twocolumn,floats,showpacs]{revtex4}
\usepackage{graphicx}
\begin{document}
\title{Shot noise near the unitary limit of a Kondo quantum dot}
\author{A. Golub}

\affiliation{ Department of Physics, Ben-Gurion University of the
Negev, Beer-Sheva, Israel \\   }
\begin{abstract}
We examine the statistics of current fluctuations  in a junction
with a quantum dot described by Kondo Hamiltonian. With the help
of Keldysh technique we calculate shot noise as a function of
applied voltage at zero temperature using the fix-point
Hamiltonian. These calculations are complementary to similar ones
performed earlier (\cite{ us}) in the mean field slave boson
 approximation. The results may be relevant for measurement of effective
 charge carriers.
\end{abstract}

\maketitle

Current noise
 has been studied
extensively in the context of mesoscopic systems in the last
couple of decades \cite{but}. The zero-frequency noise out of
equilibrium (shot noise) can yield information on charge
fluctuations in such systems. We concentrate in the following on
$T=0$,  where the thermal noise vanishes and the only contribution
to the current noise is shot noise. Due to the lack of a single
accurate method that can describe the Kondo system out of
equilibrium in all regimes, different methods need to span all
physically relevant regimes \cite{us}. Here we consider the regime
around the strongly interacting fix point of Kondo Hamiltonian
where shot-noise measurements can provide information about the
charge of the carriers. The mean field slave boson approximation
which is usually applied to describe this regime \cite{us,aguado,
hew} involves numerical calculations and it yields to the more
precise perturbation theory expansion in the least irrelevant
perturbations.

We start with the Kondo Hamiltonian for quantum
 dot in the junction $H=H_{L}+H_{R}+H_{J}$ where
\begin{eqnarray}
H_{J}&=&\sum_{\alpha\alpha'\sigma\sigma'}J_{\alpha
\alpha'}c^{\dagger}_{\alpha\sigma}(0)(\frac{1}{4}\delta_{\sigma,\sigma'}+
\vec{S}\vec{\emph{s}}_{\sigma\sigma'})c_{\alpha'\sigma'}(0)
\label{H2}
\end{eqnarray}

The first two terms correspond to non-interacting electrons in the
two leads\begin{equation}\label{lead}
    H_{L(R)}=\sum_{k,\sigma}\xi_{L(R)k}c^{\dagger}_{L(R)\sigma,k}c_{L(R)\sigma,k}
\end{equation}
 where $c_{\alpha\sigma,k}$, $ \xi_{\alpha k}$
are the electron field operator  and the electron energy of a
lead. Index $\alpha=L,R$ indicates left (right) lead. We assume
that the leads are dc-biased by applied voltage $V$. Here
$\emph{s}$ is the one half spin matrix which acts on spin index of
electron operators and $S$ is the spin operator of the dot. The
potential scattering is represented by the first term in the
brackets. We consider the case with the same dispersion law in
both leads. Near the unitary limit we follow \cite{glazman} and
map the transport through the dot into a scattering problem.  For
this purpose, it is convenient to use the basis of s and p
scattering states rather than that of the left-lead and right-lead
states. At $V=0$ the p states are decoupled from the dot. For
non-zero voltage the p-states are not decoupled, however, in the
limit of low voltages $V<< T_K$  the s, p basis is important.
Indeed, in this regime the dot is described by strongly
interacting fix-point with many body state $(b)$. Before
scattering the phases of both states, $b$ and s -state, coincide,
while passing the scattering region $b$ state acquires only an
extra phase $\pi$ compared to that of s-state. Then the fix-point
Hamiltonian can be written \cite{glazman, pust, aff} in the new
basis $(b_{\sigma,k}, a_{\sigma,k})$ (here $a$ stands for
p-states) as $H=H_0 +H_s+H_{int}$
\begin{eqnarray}
H_0&=&\sum_{k,\sigma}\xi_{k}(b^{\dagger}_{\sigma,k}b_{\sigma,k}
a^{\dagger}_{\sigma,k}a_{\sigma,k})\nonumber\\
&+&\frac{V}{2}\sum_{k,\sigma}(b^{\dagger}_{\sigma,k}a_{\sigma,k}+
a^{\dagger}_{\sigma,k}b_{\sigma,k})\nonumber\\
H_s&=&\frac{a}{\nu
T_K}\sum_{k,k',\sigma}(\xi_{k}+\xi_{k'})b^{\dagger}_{\sigma,k}b_{\sigma,k'}\label{hs}\\
H_{int}&=&\frac{b}{\nu^2
T_K}b^{\dagger}_{\uparrow}b_{\uparrow}b^{\dagger}_{\downarrow}b_{\downarrow}\label{hi}
\end{eqnarray}
where $b=2a$ and $a=1/(2\pi)$ \cite{pust} and
$b_{\uparrow}=\sum_{k}b_{\uparrow,k}$. The Kondo temperature $T_K
$ is the only energy scale of the fixed-point Hamiltonian.

 Current noise is defined as
\begin{equation}
S(t)\equiv<I(t)I(0)>-<I>^2, \label{S}
\end{equation}
where the current operator requires a form
\begin{eqnarray}
I&=&\frac{2e^2 V}{h}-\frac{ie}{\hbar}\frac{a}{2\nu
T_K}\sum_{k,k',\sigma}(\xi_{k}+\xi_{k'})\nonumber\\
&&(b^{\dagger}_{\sigma,k'}a_{\sigma,k}-
a^{\dagger}_{\sigma,k}b_{\sigma,k'})\nonumber\\
&+&\frac{ie}{\hbar}\frac{b}{2\nu^2
T_K}\sum_{\sigma}(b^{\dagger}_{\sigma}a_{\sigma}-
a^{\dagger}_{\sigma}b_{\sigma})n_{\bar{\sigma}} \label{cur}
 \end{eqnarray}
 and $n_{\sigma}=b^{\dagger}_{\sigma}b_{\sigma}$,
 $\bar{\sigma}=-\sigma $.

The noise power can be represented as a sum of two terms
 $S=S_{scat}+S_{int}$ related, respectively , to the scattering
 part of the Hamiltonian $H_{s}$ and to the interaction part $H_{int}$.

\begin{eqnarray}
S_{scat}(t,t')&=&\frac{e^2}{\hbar}\frac{a^2}{4\nu^2
T^{2}_K}\sum_{k_i,\sigma_i}(\xi_{k}+\xi_{k_1})(\xi_{k'}\nonumber\\
&&+\xi_{k'_1})<(b^{\dagger}_{\sigma,k_1}a_{\sigma,k}
-a^{\dagger}_{\sigma,k}b_{\sigma,k_1})_t\nonumber\\
&&(b^{\dagger}_{\sigma_1,k'_1} a_{\sigma_1,k'}-
a^{\dagger}_{\sigma_1,k'}b_{\sigma_1,k'_1})_{t'}>
\label{sc}\\
S_{int}(t,t')&=&\frac{e^2}{\hbar}\frac{b^2}{4\nu^4
T^{2}_K}<(b^{\dagger}_{\sigma}a_{\sigma}-
a^{\dagger}_{\sigma}b_{\sigma})n_{\bar{\sigma}}|_t(b^{\dagger}_{\sigma_1}a_{\sigma_1}\nonumber\\
&-&a^{\dagger}_{\sigma_1}b_{\sigma_1})n_{\bar{\sigma_1}}|_{t'}>
\label{si}
\end{eqnarray}
Here subscripts $t,t'$  represent time dependence of operators. We
can write the right hand side of $S_{scat}$ and $S_{int}$  as a
time ordering product in Keldysh space. (This can be simply done
by assigning Keldysh indices $2$ and $1$ to operators that depends
on $t,t'$, correspondingly. Then it is possible to express the
result in terms of Green's functions of $H_0$ Hamiltonian. Let us
perform  the Fourier transform of equations( \ref{sc},\ref{si}).
Then for zero frequency noise (after summing over spin
projections) we immediately get
\begin{eqnarray}
 S_{scat}&=&\frac{e^2}{\hbar}\frac{a^2}{\nu^2
 T^{2}_K}\int\frac{d\omega}{2\pi}\sum_{k,k_1}(\xi_{k}+\xi_{k_1})^2 [g_d ^{21}(k_1 \omega)g_d ^{12}(k
 \omega) \nonumber\\
 &-&g_{off} ^{21}(k_1 \omega)g_{off} ^{12}(k \omega)] \\
S_{int}&=&\frac{e^2}{\hbar}\frac{b^2}{
 T^{2}_K}\int\frac{d\omega_1 d\omega_2 d\omega_3}{(2\pi)^3}[(g_d ^{21}(\omega_1)g_d ^{12}(
 \omega_2)-\nonumber\\
 &&g_{off} ^{21}( \omega_1)g_{off} ^{12}(\omega_2))g_d ^{21}(\omega_3)g_d ^{12}(
 \omega_1-\omega_2+\omega_3)-\nonumber\\
&&\frac{1}{2}(g_{off} ^{21}(\omega_1)g_{d} ^{12}(
 \omega_2)-g_{d} ^{21}( \omega_1)g_{off} ^{12}(\omega_2))\nonumber\\
 &&(g_{d} ^{21}(\omega_3)g_{off} ^{12}(
 \omega_1-\omega_2+\omega_3)\nonumber\\
 &-&g_{off} ^{21}( \omega_3)g_{d} ^{12}(\omega_1-\omega_2+\omega_3))
\end{eqnarray}
  Green functions in the space of $a,b$ states can be written as
a matrix $g$ with diagonal and off-diagonal entries each being
Keldysh matrices.

\begin{equation}\label{gf}
    g_d ^{ij}(t,t')=-i<Tb^i(t)b^{\dagger j}(t')>
\end{equation}
\begin{equation}\label{gf}
    g_{off}^{ij}(t,t')=-i<Tb^i(t)a^{\dagger j}(t')>
\end{equation}
 Explicitly, they acquire a form
\begin{eqnarray}
g ^{12}(k\omega)&=&-f(\xi_{k})(g ^{R}(k \omega)-g
^{A}(k\omega))\\
g ^{21}(k\omega)&=&(1-f(\xi_{k}))(g ^{R}(k \omega)-g
^{A}(k\omega))\\
g ^{R}(k \omega)&-&g ^{A}(k\omega)= -\pi i
[\delta(\omega-\xi_{k}-eV/2)+\nonumber\\
&&\delta(\omega-\xi_{k}+eV/2)]\hat{I}+\nonumber\\
&&[\delta(\omega-\xi_{k}-eV/2)-\delta(\omega-\xi_{k}+eV/2)]\hat{\tau_x}\nonumber\\
g ^{i,j}(\omega)&=&\int d\xi_{k}g ^{i,j}(k\omega)\\
g_{off} ^{12}(\omega)&=&g_{off} ^{21}(\omega)=\pi i
[f(\omega-eV/2)-f(\omega+eV/2)]\nonumber\\
g_{d} ^{12}(\omega)&=&g_{d} ^{21}(\omega)+2\pi i \nonumber\\
&&=\pi i [f(\omega-eV/2)+f(\omega+eV/2)]\nonumber
\end{eqnarray}

With the help of these functions we get:
\begin{eqnarray}
 S_{scat}&=&\frac{e^2}{h}\frac{a^2 8\pi^2}{
 T^{2}_K}\int d\omega \omega^2 [f(\omega-\frac{eV}{2})(1-f(\omega+\frac{eV}{2}))\nonumber\\
 &+&f(\omega+\frac{eV}{2})(1-f(\omega-\frac{eV}{2}))=\nonumber\\
&&=\frac{2e^2}{h}(\frac{a^2 \pi^2}{3}) V \frac{(eV)^2}{T^{2}_K}\\
  S_{int}&=&\frac{e^2}{\hbar}\frac{\pi b^2}{
 8T^{2}_K}\int d\omega_1 d\omega_2 d\omega_3[(2-F_+(\omega_1 ))F_ +(\omega_2)\nonumber\\
 &+&F_-(\omega_1 )F_-(\omega_2 ))(2-F_+(\omega_3
 ))F_+(\omega_1-\omega_2+\omega_3)+\nonumber\\
 && \frac{1}{2}((2-F_+(\omega_1 ))F_-(\omega_2
 )+F_-(\omega_1 )F_+(\omega_2 ))((2\nonumber\\
 &-&F_+(\omega_3
 ))F_-(\omega_1-\omega_2+\omega_3)+\nonumber\\
 &&F_-(\omega_3
 )F_+(\omega_1-\omega_2+\omega_3))]\nonumber
\end{eqnarray}
where $F_\pm(\omega )=f(\omega-eV/2)\pm f(\omega+eV/2)$.

 Performing the triple integration (at $T\rightarrow 0$)
we obtain  part of the noise $S_{int}$ which is induced by
interaction Hamiltonian $H_{int}$. Thus the total shot noise power
in the limit of small voltages $eV<<T_K $ acquires a form

\begin{equation}\label{int}
    S_{int}=\frac{e^2V}{h}\pi^2(\frac{3}{2}b^2  +
    \frac{2a^2 }{3})\frac{(eV)^2}{T^{2}_K}
\end{equation}
 For completeness  we also calculate nonlinear conductance
 \cite{glazman,pust}. Similarly to the noise  the current consist
 of to parts
 $j=j_{scat} +j_{int}$, where $j_{scat}$ is the
scattering part
\begin{equation}\label{j1}
    j_{scat}=\frac{e}{\hbar}<\hat{T}[H_s,N_R]^{1}_t \int dt'\sigma_z^i H_s ^i (t')>
\end{equation}
and $j_{int}$ interacting part
 \begin{equation}\label{j2}
      j_{int}=\frac{e}{\hbar}<\hat{T}[H_{int},N_R]^1 _t \int dt'\sigma_z^i H_{int} ^i (t')>
\end{equation}
Here $\hat{T} $ represents time-ordering operator. We also
explicitly
 introduced Keldysh indices $i=1,2$ and defined
 $\sigma^1=1,\sigma^2=-1$.

 Thus for the scattering part of the current we obtain
\begin{eqnarray}\label{j1}
    j_{scat}&=&\frac{e}{\hbar}\frac{a^2}{\nu^2
 T^{2}_K}\int\frac{d\omega}{2\pi}\sum_{k,k_1}(\xi_{k}+\xi_{k_1})^2 (f(\xi_{k})\nonumber\\
 &-&f(\xi_{k_1}))g_d ^{dif}(k_1 \omega)g_{off} ^{dif}(k
 \omega)
\end{eqnarray}
where $g^{dif}=g^R- g^A$ . The frequency integration and momentum
summation can be easily done with the result:
\begin{equation}\label{rj1}
    j_{scat}=\frac{2e^2}{h}\frac{a^2\pi^2}{3}(\frac{eV}{T_K })^2 V
\end{equation}
For interacting part we have
\begin{equation}\label{j2r}
 j_{int}= \frac{2e^2}{h}\frac{b^2\pi^2}{3}\frac{5}{4}(\frac{eV}{T_K })^2 V
\end{equation}
Thus total nonlinear conductance at zero temperature and in the
case of small voltages $eV<<T_K $ acquires a form
\begin{equation}\label{con}
    \sigma=\frac{2e^2}{h}[1- \pi^2(a^2+b^2 \frac{5}{4})(\frac{eV}{T_K
    })^2]
\end{equation}
This result is in agreement with that obtained in \cite{pust}.
\begin{acknowledgments}
This work was  initiated by
 the problem of  effective charge of carriers, the
problem which was put forward by Y. Oreg and with whom (and also
with E. Sela) the author has been communicated. Their interests in
the noise calculations at the unitary limit of Kondo dot has
stimulated this work. I would like to thank them both for
discussions.

\end{acknowledgments}

\end{document}